

\font\sixrm=cmr6
\font\sixi=cmmi6
\font\sixsy=cmsy6

\font\sevenrm=cmr7
\font\seveni=cmmi7
\font\sevensy=cmsy7

\font\tenrm=cmr10

\font\twelverm=cmr12
\font\twelvei=cmmi12
\font\twelvesy=cmsy10 at 12pt
\font\twelveit=cmti12
\font\twelvesl=cmsl12
\font\twelvebf=cmbx12
\font\twelvett=cmtt10 at 12pt

\font\fourteenrm=cmr10 scaled \magstep3
\font\fourteenbf=cmbx10 scaled \magstep3

\def\twelvepoint{%
\def\rm{\fam0\twelverm}%
\def\it{\fam\itfam\twelveit}%
\def\sl{\fam\slfam\twelvesl}%
\def\bf{\fam\bffam\twelvebf}%
\def\tt{\fam\ttfam\twelvett}%
\def\cal{\twelvesy}%
 \textfont0=\twelverm
  \scriptfont0=\sevenrm
  \scriptscriptfont0=\sixrm
 \textfont1=\twelvei
  \scriptfont1=\seveni
  \scriptscriptfont1=\sixi
 \textfont2=\twelvesy
  \scriptfont2=\sevensy
  \scriptscriptfont2=\sixsy
 \textfont3=\tenex
  \scriptfont3=\tenex
  \scriptscriptfont3=\tenex
 \textfont\itfam=\twelveit
 \textfont\slfam=\twelvesl
 \textfont\bffam=\twelvebf
 \textfont\ttfam=\twelvett
 \baselineskip=15pt
}


\hsize     = 152.0mm
\vsize     = 225.5mm 
\hoffset   =   0.0in
\voffset   =   0.0in
\topskip   =  15pt
\parskip   =   0pt
\parindent =  21pt
\nopagenumbers

\newskip\one
\one=15pt
\def\One{\vskip-\lastskip\vskip\one}

\newcount\LastMac
\def\Skipe{1}  
\def\Hae{3}    
\def\Hbe{4}    
\def\Hce{5}    

\def\SkipToFirstLine{
 \LastMac=\Skipe
 \dimen255=150pt
 \advance\dimen255 by -\pagetotal
 \vskip\dimen255
}

\def\Raggedright{%
 \rightskip=0pt plus \hsize
 \spaceskip=.3333em
 \xspaceskip=.5em
}

\def\Fullout{
 \rightskip=0pt
 \spaceskip=0pt
 \xspaceskip=0pt
}
%
%

\newbox\grsign \setbox\grsign=\hbox{$>$}
\newdimen\grdimen \grdimen=\ht\grsign
\newbox\laxbox \newbox\gaxbox
\setbox\gaxbox=\hbox{\raise.5ex\hbox{$>$}\llap
     {\lower.5ex\hbox{$\sim$}}}\ht1=\grdimen\dp1=0pt
\setbox\laxbox=\hbox{\raise.5ex\hbox{$<$}\llap
     {\lower.5ex\hbox{$\sim$}}}\ht2=\grdimen\dp2=0pt


\def\ct#1\par{
 \One
 \Raggedright
 \fourteenbf\baselineskip=16pt
 \noindent\centerline{
 #1}
}

\def\ca#1\par{
 \One
 \Raggedright
 \fourteenrm\baselineskip=16pt
 \noindent
 \centerline{#1}
 \One
}

\def\aa#1\par{
 \Raggedright
 \twelverm\baselineskip=14.4pt
 \noindent
 \centerline{#1}
 \hfil\break
}

\def\abstract#1\par{
        \centerline {\twelvebf ABSTRACT}
 \One
 \ifnum\LastMac=\Skipe \else \One\fi
 \LastMac=\Hae
 \Fullout
 \tenrm\baselineskip=12pt
 \vbox{
 \leftskip =21pt
 \rightskip=21pt
 #1
}
 \One
}

\def\ha#1\par{
 \ifnum\LastMac=\Skipe \else \One\fi
 \LastMac=\Hae
 \Raggedright
 \twelvebf\baselineskip=15pt
 \noindent
 \uppercase{#1}
 \twelvepoint\rm
 \One
 \Fullout
}

\def\hb#1\par{
 \LastMac=\Hbe
 \One
 \Raggedright
 \twelvebf\baselineskip=15pt
 \noindent
 #1
 \twelvepoint\rm
 \One
 \Fullout
}

\def\hc#1\par{
 \LastMac=\Hce
 \One
 \Raggedright
 \twelvebf\baselineskip=15pt
 \noindent
 #1
 \twelvepoint\rm
 \One
 \Fullout
}

\def\ref{
 \Fullout
 \twelvepoint\rm
  \noindent
  \hangindent=21pt
  \hangafter=1
}

\def\toprightinsert #1#2#3#4{
 \topinsert
 \vbox to #2truein{
 \leftskip=#3truein
 \vskip #1truein
 \raggedright\parindent=21pt
 #4
 \hfil\break\vfil
 }
 \endinsert
}

\message{Font definitions, page makeup parameters, new commands,}
\message{and design element definitions complete; leaving Specs.tex}

\centerline{\it
To appear in the proceedings of the conference on ``Mass Transfer Induced
Activities in Galaxies''}
\centerline{\it held in Lexington, Kentucky, Apr. 1993, I. Shlosman
(ed.)\footnote*{\it Available at Preprint Server: astro-ph@babbage.sissa.it,
get astro-ph/9308030}}

\ct The Galactic Center -- an AGN on a starvation diet\par

\ca Heino Falcke and Peter L. Biermann\par

\aa MPIfR, Auf dem H\"ugel 69, D-53121 Bonn, Germany,
(HFalcke@mpifr-bonn.mpg.de)

\smallskip

\abstract
The Galactic Center shows evidence for the presence of three important
AGN ingredients: a Black Hole ($M_\bullet\sim10^6M_\odot$), an
accretion disk ($10^{-8.5} - 10^{-7} M_\odot/{\rm yr}$) and a powerful
jet (jet power $\ge$ 10\% disk luminosity). However, the degree of
activity is very low and can barely account for the energetics of the
whole central region. Neverthelss, in the very inner arsecond the
central engine becomes dominant and provides an interesting laboratory
f[222zor the physics of central engines (Black Holes) in galactic nuclei.
We therefore give an overall picture of the central arcsecond where we
link the radio emission and the heating of the ambient medium to a
weakly accreting disk surrounding a massive Black Hole.

\par

\ha 1 Introduction\par
The dynamical center of the Galaxy is the radio point source Sgr A*,
which is also the center of the central star cluster (Eckart et al.
1993). Investigations of the enclosed mass in the central region show
that there is evidence for a mass concentration of the order of
10$^6$M$_{\odot}$ within the central arcsecond (Genzel and Townes
1987).  There is good reason to assume that this ``dark mass'' indeed
is the mass of a massive Black Hole (BH) powering Sgr A*. The total
spectrum of this source from radio to NIR was compiled by Zylka et al.
(1992).  There is a flat radio spectrum up to 7mm, and a steeply
rising submm spectrum, which Zylka et al. interpret as thermal
emission from a dust torus surrounding the BH. In the FIR one finds a
spectral break at 30$\mu$m indicated by upper limits and a third
spectral component rising in the NIR, which has been interpreted as
emission from an accretion disk around the BH.\par

\ha 2 Hertzsprung-Russell diagram for the Sgr A* disk\par

Because of strong obscuration in the galactic plane we probably will
never be able to measure exactly the optical and UV part of Sgr A*,
which is needed to discriminate between different disk models.
Nevertheless, there are a number of parameteres we can infer by
indirect means to constrain possible models. For example, we know that
in the outer part of the Galactic Center region almost 50\% of the
starlight is absorbed in the interstellar dust (Cox \& Mezger 1989).
IR measurements show that the dust concentration strongly peaks around
Sgr A*, therefore we assume that all the luminosity of Sgr A* beyond
the NIR is absorbed in the dust and reradiated at longer wavelength.
Zylka et al. estimated the total luminosity in the central 30'' to be
$1.5\cdot10^6 L_\odot$.  Interestingly the total luminosity of the
stellar star cluster, as extrapolated from the outer parts (Falcke et
al. 1993a) is of comparable order, thus one obtains an upper limit for
the disk luminosity of $L_{\rm disk}<7\cdot 10^5 L_\odot$.  Further
limits for $L_{\rm disk}$ and the effective temperature $T_{\rm eff}$
can be found by assuming that the NIR measurements represent the
Rayleigh-Jeans tail of a Black-Body, leading to $L_{\rm disk}>7\cdot
10^4 L_{\rm disk}$ and $20,000 {\rm K} < T_{\rm eff}< 40,000 {\rm
K}$.\par
\toprightinsert{0}{4.7}{0}
{Fig.~1---{\tenrm\baselineskip=12pt\Fullout HR diagrams for BHs. The
figure on the left gives an overwiev for the whole parameter range of
astrophysically relevant BHs. The small areas cover all possible
values for effective temperature and luminosity of a BH of given mass
and accretion rate but different inclination angles and angular
momenta $a$. Areas at the same horizontal line have the same absolute
accretion rate, areas at the same diagonal line from lower left to
upper right have the same mass and areas at the same diagonal from the
lower right to the upper left have the same accretion rate in terms of
their Eddington rate. The boxes denote schematically the position of
AGN, Sgr A* and stellar mass BHs. The figure on the right is a zoomed
 version
of the HRD, however, for an edge-on disk and a fixed BH mass
of $2\cdot10^6M_\odot$.}}

If we construct a HR-diagram for BHs (Falcke et al. 1993a), we find
that for a given BH mass of $2\cdot 10^6 M_\odot$ the data is
consistent with a maximally rotating BH accreting $10^{-7}$ -
$10^{-8.5}M_\odot/{\rm yr}$ in a disk seen edge on (Fig.  1). A
lighter BH with $M_\bullet=10^3 M_\odot$ does not fit the current data
at all. The accretion rate we find is more than 5 orders of magnitude
lower than the Eddington limit of the BH and 7 orders of magnitude
lower than in average AGN -- the Galactic Center resembles an AGN
on a starvation diet.

\eject
\ha 3 Black Hole H{ II} Region\par
\toprightinsert{0}{3.9}{0}
{Fig.~2--- {\tenrm\baselineskip=12pt\Fullout The temperatures of a
test cloud of constant density ($n=10^4/{\rm cm}^{3}$) exposed to a disk
spectrum for an accretion rate of $10^{-7} M_\odot/{\rm yr}$ around
maximally rotating BHs of mass $10^6 M_\odot$ (left) and $10^3
M_\odot$ (right). The disk lies horizontally in the center and is seen
edge on.}}

To learn more about the hidden UV spectrum of Sgr A*, one has to apply
indirect methods, i.e. the heating and ionization of ambient gas by
the disk spectrum. From the HR diagram one can see, that the disk is
at the edge of producing an appreciable amount of ionizing UV photons,
therefore the state of the ambient gas will be very sensitive towards
changes of the parameters in the BH/accretion disk system. To test
this we took the photoionization code HOTGAS developed by
Schmutzler \& Tscharnuter (1993) to calculate gas temperatures of a
test cloud with constant density at different spatial positions around
Sgr A* neglecting all dynamical and optical depth effects (Fig. 2).

For a set of parameters with $M_\bullet=10^6 M_\odot$, $\dot M=10^{-7}
M_\odot/{\rm yr}$ and a maximally rotating BH, we find a very
anisotropic temperature distribution, which reflects the relativistic
beaming of the disk spectrum at high inclination angles, heating the
gas to temperatures of $T_{\rm gas}\ge10.000$ K. A moderate change of
the angular momentum of the BH from $a=0.9981$ to $a=0.9$ would be
enough to suppress the heating almost completely. On the other hand, a
low mass BH (e.g. $10^3 M_\odot$) inevitably leads to much
higher temperatures ($10^5-10^6$ K) in the inner arcsec of the
Galactic Center. No configuration is able to heat the total H II
region Sgr A*. Nevertheless, once we are able to determine temperature
and density distribution of the gas in the inner arcsecond of the GC
we will have a powerful tool to discriminate between different
models.

\eject
\ha 4 Sgr A*: a jet?\par

After having shown that the IR-NIR spectrum is consistent with the
presence of an accretion disk, it is straightforward to
postulate the existence of a radio jet created at the inner edge of
the disk producing the compact flat spectrum radio emission -- a
feature seemingly related to disks. This jet/disk link directly
imposes an important constraint: The jet can not carry away
more matter and more energy than is provided by the accretion process.

We adopt the Blandford \& K\"onigl (1979) jet model, which
assumes a conically expanding, supersonic jet, with constant velocity
and an internal gas pressure  dominated by a turbulent magnetic
field being in energy equipartition with relativistic electrons.

To account for the jet/disk link, we express the mass loss due to
the jet $\dot M_{\rm jet}$ and the total energy of the jet
$Q_{\rm jet}$ in terms of the disk accretion rate $\dot
M_{\rm disk}$, such that

$$Q_{\rm jet}=q_{\rm j}\dot{M}_{\rm disk} c^2\;\;\;\;{\rm and}\;\;\;\;
\dot M_{\rm jet}=q_{\rm m} \dot{M}_{\rm disk}.\eqno(1) $$

Using the above parameterization, we find for the flux of Sgr A* a
flat spectrum with $F_\nu={\rm const}$ and an absolute value of

$$ F_\nu=1\,\,{\rm Jy}\cdot\,{\cal D}(i,\gamma_{\rm
j})^{13/6}{{\sin { i^{{1/ 6}}}}}\, \left({{\cal M}\over 3
}\right)^{-11/6}
{{\left({\Lambda\over9}\right)}^{-{5/6}}} \left(\gamma_{\rm
j}\beta_{\rm j}\,{q_{\rm m}\over 3\%}\,{\dot{M}_{\rm
disk}\over 10^{-7}M_{\odot}/{\rm yr}}\right)^{17/12}
\eqno(2)$$
Here $\cal D$ is the Doppler factor, $i$ the inclination of the jet
axis, $\cal M$ the Machnumber of the jet, $\beta_{\rm j}$ the velocity
of the jet, $\gamma_{\rm j}$ the relativistic $\gamma$ factor of the
jet and $\Lambda$ a logarithmic correction factor describing the
relativistic electrons.

We see that even with the extremely low accretion rate in Sgr A* it is
possible to feed a radio jet producing 1 Jy emission with a reasonable
set of parameters. However, an appreciable fraction of the total mass
accretion rate ($\ge3\%$) has to be expelled by the jet. Using the
energy equation of this system (Falcke et al 1993b) this translates to
a jet power of $Q_{\rm j}\ge4\% \dot M_{\rm disk}c^2$. Thus the ratio
of the total jet power to the disk luminosity $L_{\rm disk}\le 30 \%
\dot M_{\rm disk} c^2$ has a lower limit of $Q_{\rm jet}/L_{\rm
disk}\ge12\%$.

Alas, there is a problem: for 20 years now, VLBI radio observations
tell us that Sgr A* is a point source at all wavelengths. How can this
be a jet? An answer is found by examining the structural
predictions of the jet/disk model. The physical scale of the
synchroton emission depends inversely on frequency yielding for the
Sgr A* set of parameters


$$ z\simeq2\cdot 10^{13} {\rm cm} \left({{43 GHz}\over{\nu}}\right)
\left({\gamma_{\rm j}\beta_{\rm j}
\over\sqrt{1+\left(\gamma_{\rm j}\beta_{\rm j}\right)^2}}
\sqrt{{9\over\Lambda} {3\over{\cal M}}} \,{q_{\rm m}\over
3\%}\,{{\dot{M}_{\rm disk}}\over{10^{-7}M_{\odot}/{\rm
yr}}}\right)^{2/3},
\eqno(3)$$

For the low accretion rate of Sgr A*, the scale of the jet then
is smaller than the resolution of VLBI even at 43 GHz.  Thus, one
should see at best a marginally resolved central core. And indeed,
this is confirmed by recent 43 GHz VLBI observations of Sgr A*
(Krichbaum et al. 1993), where the emission ist still dominated by an
unresolved central core which, however, is slightly elongated
suggesting an underlying jet structure.

There is another interesting obervational consequence associated with
equation (3). We can turn the argument around and ask: What is the
shortest possible wavelength $\lambda_{\rm break}$ emitted by such a
jet, namely the one emitted at the shortest length scale.  For a given
central mass, the smallest possible scale is the scale of a BH which
is $R_{\rm g}=1.5\cdot 10^{11} M_\bullet/10^6 M_\odot$ cm. A
reasonable and conservative guess for the smallest jet scale then also
would be of the order of several $R_{\rm g}$, say $z_{\rm min}\ge
10\cdot R_{\rm g}$.  For a $10^6 M_\odot$ BH we obtain $\lambda_{\rm
break}\le560 \mu$m and for a $10^3M_\odot$ BH we obtain $\lambda_{\rm
break}\le.56\mu$m.  For wavelength shorter than $\lambda_{\rm break}$
we would expect to see a steepening of the spectral index from 0 to
-0.5 or even steeper.

This explains very well the oberved lack of far infrared emission
(shortwards 30$\mu$m) from Sgr A* if there is indeed a $10^6 M_\odot$
BH. On the other hand, there is no such argument for a $10^3 M_\odot$
BH. As the break frequency should in this case be somewhere in the NIR
we would rather expect a continuing flat spectrum visible also in the
FIR, which is not observed.

\ha 4 Conclusions\par
Spectral and structural information of Sgr A* are consistent with the
standard AGN triad Black Hole, jet and accretion disk, with converging
evidence for a supermassive Black Hole ($10^6 M_\odot$), low accretion
rate ($10^{-8.5} - 10^{-7} M_\odot/{\rm yr}$) and a powerful radio-jet
(as compared to the disk luminosity). This set of parameters is
consistent with the NIR data, the dust luminosity, the radio spectrum,
the size of the radio source and the lack of non-thermal FIR emission.

However, one question is still unanswered: Why is the central
accretion rate so extremely low? Which mechanism prevents the large
amount of gas in this region from being accreted onto the BH?

A plausible explanation would be to assume that the accretion process
varies strongly radially and in time. Thus the central accretion rate
could have been much higher in earlier epochs.  A dust torus in the
inner arcsecond -- as suggested by submm observations -- could be just
the outer part of a non-stationary accretion disk, serving as a
reservoir where matter is temporarily stored until stronger accretion
process in the inner parts sets in again.

\smallskip

{\it Acknowledgement:} We want to thank P. Mezger, W. Duschl, T.
Krichbaum, K. Mannheim, M.+G. Rieke and R. Zylka for extensive
discussions on this topic. HF is supported by DFG grant (Bi 191/9). We
thank the staff of the Vatican Observatory -- where this article was
written -- for their kind hospitality.
\eject

\ha References \par
\ref Blandford, R.D., K\"onigl, A. 1979, {\it ApJ}, {\bf 232}, 34\par
\ref Cox, P., Meyger, P.G. 1989, {\it A\&AR}, {\bf 1}, 49\par
\ref Eckart, A., Genzel, R., Hofmann, R. et al. 1993, {\it ApJ}, {\bf 407},
 L77\par
\ref Falcke, H., Biermann, P.L., Duschl, W.J., Mezger, P.G. 1993,
{\it A\&A}, {\bf 270}, 102\par
\ref Falcke, H., Mannheim, K., Biermann, P.L 1993b, {\it A\&A}, in press\par
\ref Krichbaum T.P., Zensus J.A., Witzel A., Mezger P.G.,
 Standke K. et al., 1993, {\it A\&A},{\bf  274}, L37 \par
\ref Schmutzler, T., Tscharnuter,  1993, {\it A\&A}, {\bf 273}, 318\par
\ref Zylka, R., Mezger, P.G., Lesch, H. 1992, {\it A\&A}, {\bf 261}, 119
\vfill\supereject
\message{TTTTTTTTTTttttttttttthhhhhhhhhhaaaaaaaaattttttt's all Folks!}
\bye